# Inventions on reducing keyboard size
## A TRIZ based analysis

**Umakant Mishra**

Bangalore, India
umakant@trizsite.tk
http://umakant.trizsite.tk

**Contents**

# 1. Introduction

A conventional computer keyboard consists of as many as 101 keys. The keyboard has several sections, such as text entry section, navigation section, numeric keypad etc. The text entry section contains the standard character keys, navigation section contains cursor movement and page control keys, numeric keypad contains numeric keys and function keys section contain function keys and special keys. Although the increased number of keys helps better interaction with the computer it unfortunately increases the size of the keyboard.

The size of the keyboard is a major inconvenience for portable computers, as they cannot be carried easily. The following are certain circumstances, which compels to reduce the size of a keyboard.

**1.1 Need for reducing the size of the keyboard**

- The portable and handheld computers are small in size and hence cannot accommodate a large keyboard inside the box.

- Large size keyboards are difficult to carry.

- Large keyboards occupy more desk space.

- The large keyboards contain large number of keys, which may be confusing for new users.

**1.2 Problems in reducing the size of keyboard**

- Reducing the size of keyboard may not allow us to keep all the keys of a standard keyboard.

- Reducing the number of keys may need to change the standard key layout that will need relearning the keyboard operation for the existing users.

- Reducing the size of the keyboard may reduce the size of keys, which will be inconvenient to operate.

- Reducing the size of the keyboard will reduce the gap between the keys, which will cause inconvenience in typing.

- Reducing the size of the keyboard may be convenient to carry but not convenient for typing.

### 1.3 TRIZ contradictions with keyboard size

**Problem**- reduction in the size of the keyboard may lead to reduction in the size of keys. Reduction in the size of keys leads to pressing multiple keys because of the natural size of the human finger. We need the keyboard to be small but the keys to be operated precisely **(contradiction).**

**Solution-** Use a pointed stick or stylus to operate the keyboard instead of operating with fingers **(Principle-24: Intermediary)**.

**Problem**- the keys in a keyboard need to have adequate space to be operated by the fingers. Under the condition, reduction in the size of a keyboard may lead to reduction in the number of keys. Reduction in the number of keys may support less number of functions. We need more number of functions, but we want only less number of keys **(Contradiction)**.

**Solution**- Use multi-stroke per character to generate more number of command functions from less number of keys **(Principle-17: Another dimension)**.

**Problem**- In a multi-stroke mechanism the user has to press two or more keys to get a character, where the user is confused about which key to press after what. We want to use multi-stroke mechanism to generate more characters, but we don't want users to be confused **(Contradiction)**.

**Solution-** Use an operating guide to assist users on selecting keystrokes **(Principle-8: Counterweight)**.

**Problem-** Reducing the size of the keyboard may be convenient to carry but not convenient for typing. We need a large keyboard for typing but a small keyboard to carry **(Contradiction)**.

**Solution-** Use a collapsible keyboard which can be made large while typing and folded while carrying **(Principle-15: Dynamize)**.

## 2. Inventions on reducing keyboard size

### 2.1 Compressible keyboard (Patent 5141343)

**Background problem**
One of the typical problems in miniaturizing the computer is the ergonomic factor of the keyboard and the size of human finger. Again most keyboards are of fixed size which does not fill to all different users. It is necessary to make the size of the keyboard flexible so as to fit to the size of the finger.

**Solution provided by the invention**
William Roylance invented a method (Patent 5141343, Issued in Aug 92) to make a keyboard protractable and compactible by changing the relative spacing between the keys. The invention comprises an extensible and contractible housing, a frame which extends and compresses with the housing, keypads, and associated switches installed in the frame and thereby moved in relationship to each other so that the spatial orientation of the switches is protractable and compactable.

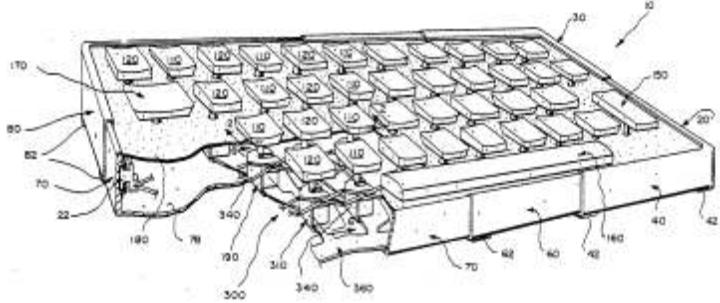

This invention of an extensible and compressible keyboard can be used for computers and other similar applications where storage space is a prime consideration. The keyboard provides adjustable spacing between the keys whereby a compact storage unit is expanded to a full-size, full function, and normally spaced standard keyboard or, as importantly, to a size which best fits the fingers of an individual user.

### TRIZ based analysis

The size of the keyboard should be big enough to allow the human fingers to operate and the size of the keyboard should be small enough to fit with the portable computer **(Contradiction)**.

This invention removes the space between the keys to reduce its size when not used **(Principle-2: Taking out)**.

The keyboard is expanded to full-size when used and compressed to small size when carried **(Principle-15: Dynamize)**.

### 2.2 Pivoting electronic keyboard keys (Patent 5329278)

### Background problem

Standard computer keyboards are too large and not suitable to carry with small handheld computers. Although there are many small keyboards, they are mostly operated by only one finger, which results in very slow operation.

There are some inventions to operate small keyboards with multiple fingers but they contain less number of keys and key layouts are very different from the conventional QWERTY keyboard, which make them mostly unacceptable.

It is necessary to keep the layout of QWERTY keyboard, (which necessitates retaining all the keys of the QWERTY keyboard), necessary to reduce the size but operate with two hands like a conventional large keyboard (Contradiction).

**Solution provided by the invention**

Michael Dombroski solved this problem by disclosing pivoted keys for a keyboard (Patent 5329278, assignee-Nil, issued July 1994). The keyboard uses key pairs, which are two side-by-side keys facing each other, mounted for pivotal movement. The key face surfaces slant towards each other and look upwardly concave. The keyboard is small in size with adequate vertical travel and minimal lateral travel.

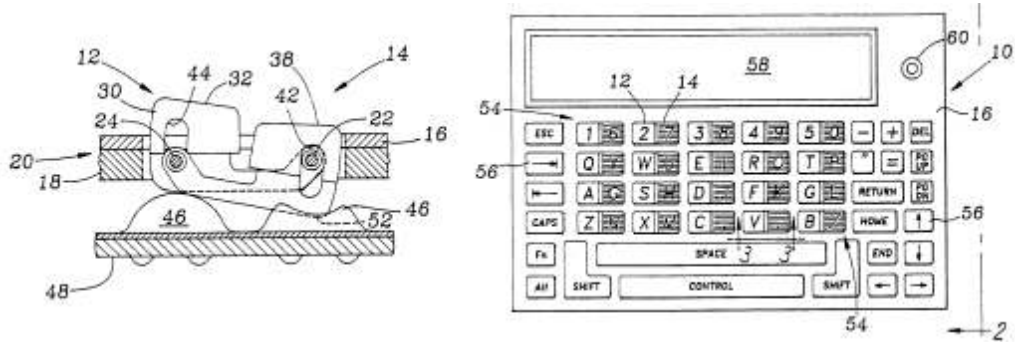

The advantage of this invention is that we can get all the keys of a standard keyboard in almost half of the space. The user need not learn typing again as he is already familiar with the standard keyboard.

**TRIZ based analysis**

The invention uses key pairs and reduces the width of the keyboard by half **(Principle-17: Another dimension)**.

The key pairs are pivotally mounted so that both the keys cannot be pressed at the same time **(Principle-9: Prior counteraction)**.

**2.3 Compact computing device having a compressible keyboard (Patent 5870034)**

**Background problem**

The small size of laptops, palmtops and PDAs need small keyboards to fit into the size of the box. But a small keyboard does not give the typing comfort of a conventional PC keyboard. Therefore sometimes the laptop users connect external keyboards to their laptops for intensive typing jobs, which take up additional space.

This situation creates a typical contradiction with the keyboard size, i.e., the keyboard should be large enough to give us typing comfort and small enough to be carried with a small laptop.

**Solution provided by the invention**

Anthony Wood disclosed a method (US patent 5870034, assigned to Texas Instruments, Feb 99) to reduce the full size keyboards to small portable computers. The size of the keyboards is increased while in use, which permits comfortable typing. The invention eliminates the unwanted space between keys during storage, which reduces its size to fit into the laptop making it easy to carry.

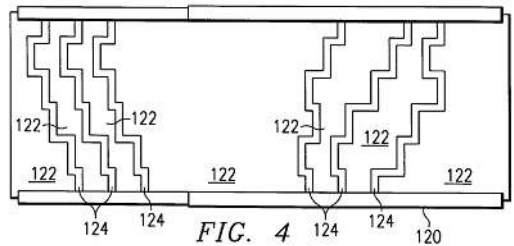

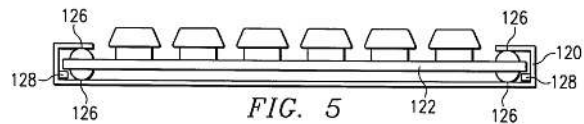

The invention discloses several sections and collapsible skirts to provide additional space for these sections. These sections are compressed in the storage position but expanded when deployed.

**TRIZ based analysis**

The keyboard should change its size (to big or small) according to our need **(Ideal Final Result)**.

**Contradiction**: We need the laptop keyboard to be large for typing comfort. But we don't want the large keyboard to carry. We need the laptop keyboard to be small to fit into the laptop and easy to carry, but it should be large like a desktop keyboard when deployed. This creates a physical contradiction.

**Solution**- we can reduce the size while carrying and increase the size while typing (solving by time).

The invention discloses a resizable keyboard that can expand and contract depending on our need **(Principle-15: Dynamize)**.

The invention eliminates the gaps between the keys when compressed and restore the gaps between the keys when expanded **(Principle-34: Discard and recover)**.

**2.4 Elevated separate external keyboard apparatus for use with portable computer (Patent 5894406)**

**Background problem**

The keyboards of the portable computers are relatively smaller than the conventional keyboards, which is not convenient for typing for long time. To overcome this problem, some laptops have keyboard ports to attach a regular keyboard for typing comfort. However this solution uses extra desk space and leads to visual and operational limitations.

**Solution provided by the invention**

Michael Blend and Allan Lichtenberg (Patent 5894406, April 99) disclosed a method of embedding a larger external keyboard on the laptop. The external keyboard is specially built to have a cavity to fit on key plate of the portable computer. In other words the external keyboard is kept on the laptop keyboard resting on it. This solves the problem of occupying extra space for the external keyboard, and solves the visual and operational problems of using an external large keyboard.

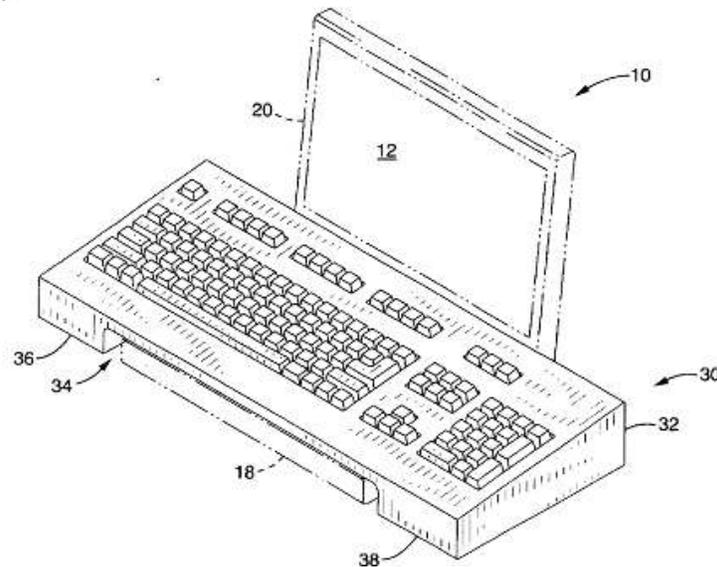

**TRIZ based analysis**

Contradiction: We need a large external keyboard to operate with a portable computer but we don't want to waste the space for two keyboards **(Contradiction)**. We want to use an external keyboard but we don't want to have excessive space between the screen and the keyboard which would cause difficulties in viewing both the monitor and keyboard. **(Contradiction)**.

The invention discloses a keyboard which sits on the laptop keyboard without wasting additional desk space **(Principle-7: Nesting)**.

**2.5 Expandable keyboard for small computers (Patent 5938353)**

**Background problem**

The small computers need small keyboards for carrying along with the computer. The user reduces his typing speed when the size of the keyboard is even slightly reduced.

One option is make a folding structure of the keyboard. The folding structure of the keyboard, although takes less surface space, increases the depth (or height) significantly. There is a need to reduce the size of the keyboard without increasing its height unlike a folding keyboard.

**Solution provided by the invention**

Butler invented an expandable keyboard (Patent 5938353, Aug 99) for small computers whose keys have interlocking side edges and are mounted on a scissor-linkage that allows said keyboard to contract to a width that is much narrower than the width of a standard desktop computer keyboard.

The invention is based on the truth that a significant amount of space exists between the raised side edges of any pair of adjacent keys in a standard (19 mm) pitch. This gap between the keys is about half of the width of the key. The problem can be addressed if these gaps are available or expanded in "open" position, yet eliminated or contracted in "closed" position. The elimination of gaps between the keys can reduce a 28-29 cm keyboard to 17-18 cm length without increasing the depth.

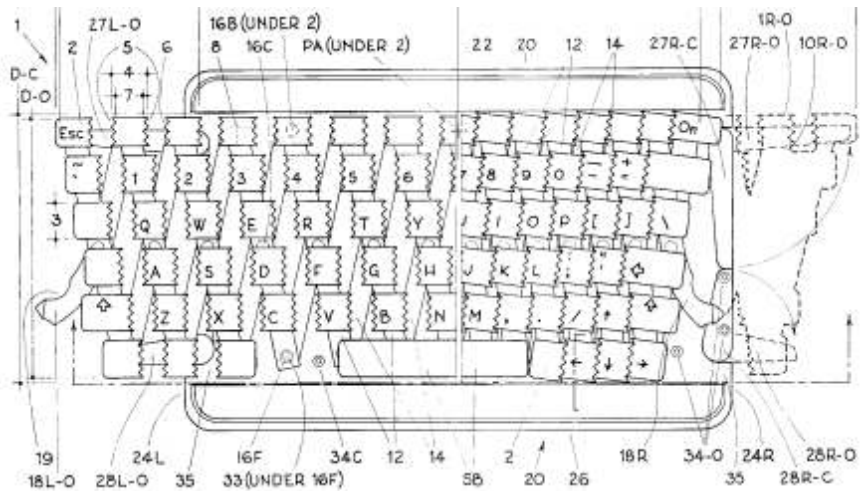

**TRIZ based analysis**

The keyboard should be small to carry and big to provide typing comfort **(Contradiction)**.

The invention eliminates the gap between keys when the keyboard is packed for carrying and expands the gap between keys when used for typing **(Principle-34: discard and recover)**.

## 2.6 Keyboard and notebook type computer (Patent 5995024)

**Background problem**

The operating surface of the laptop keyboard is very small because of the miniaturization of the computer. The low operating surface of the laptop keyboards makes them inconvenient for typing. How to increase the operating surface without increasing the size of the keyboard?

**Solution provided by the invention**

Kambayashi et al. invented a keyboard (Patent 5995024, Assigned to Fujitsu Limited, Nov 99) that uses a similar method to provide more operating space. The invention realizes that the transverse directional length is useful for typing comfort; hence we can reduce the vertical length of the keyboard without affecting the typing comfort.

According to the invention the size of the keys are reduced in the vertical direction and gaps between keys are reduced in the vertical direction without changing the size of the keys or gaps between the keys in the transverse direction. This makes the length of the keyboard larger in transverse direction than that in the vertical direction.

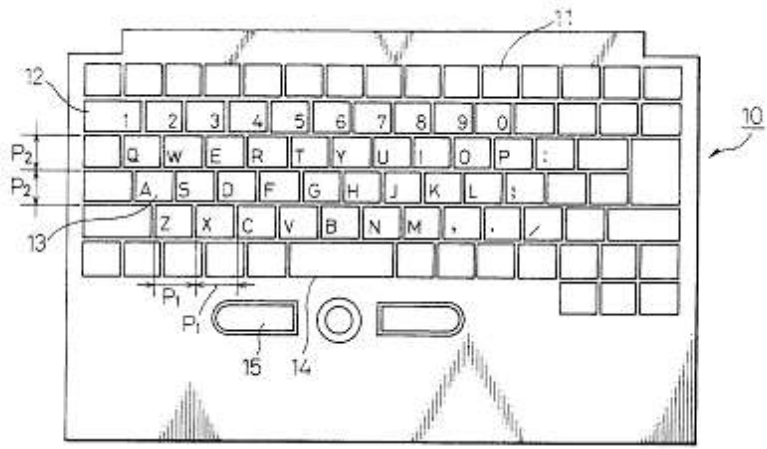

As the vertical directional length of the upper surface of the keyboard is shortened, a relatively large armrest area is obtained, improving the operability results. As the transverse directional length of the key top is longer than the vertical directional length, it provides the operational comfort and reliability.

**TRIZ based analysis**

The keyboard should provide larger operating space for typing comfort, but providing more operating space will eventually increase the size of the keyboard which will not fit into a laptop. **(Contradiction)**

The invention reduces the vertical size of the keys and vertical space between keys, which reduces the vertical size of the keyboard keeping the transverse size in tact. This provides operating space as well as more armrest area **(Principle-2: Taking out, Principle-17: Another dimension)**.

## 2.7 Special key on keyboard (Patent 6198474)

**Background**

The size of a laptop keyboard should be small to fit into the laptop box. But the size of the laptop keyboard should not sacrifice its efficiency and comfort. How do we reduce the size of the keyboard without reducing its benefits?

**Solution provided by the invention**

Roylance invented a method of reducing the size of a keyboard by reorganizing the keys on a keyboard (US patent 6198474, Mar 2001) but without disturbing the basic QWERTY layout. According to the invention the size of the keyboard is reduced by removing the key-modifier keys and assigning those functions to other keys. A special key is created by dividing the conventional long spacebar and used in concert with the alphanumeric keys of the keyboard to provide key-modifier, macro and special character functions. This invention is claimed to be quite space efficient for pocket organizer or pocket computers.

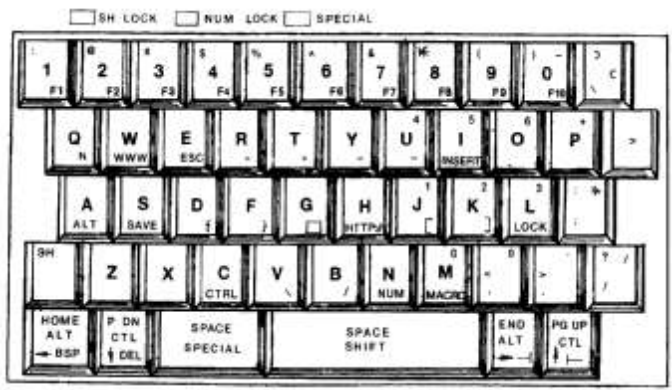

**TRIZ based analysis**

The single spacebar of a standard keyboard does not utilize the full dexterity of the human thumb. The invention splits the space bar to make two keys and use one of the spacebars as a thumb operated special key **(Principle-22: Blessings in Disguise, Principle-1: Segmentation)**.

The special key (created out of the spacebar) is used in concert with other alphanumeric keys to produce special character functions **(Principle-5: Merging)**.

## 2.8 Compact keyboard (Patent 6431776)

**Background problem**

A conventional keyboard contains a plurality of alphanumeric keys and thereby occupies more space on table because of its size. A compact keyboard is desirable to save the desk space and save raw materials to produce it.

**Solution provided by the invention**

Tzeng invented a compact keyboard (patent 6431776, assigned to Darfon Electronics, Aug 02) having a novel key arrangement, which reduces the width of the keyboard. The new keyboard contains all the keys of a standard keyboard including a numeric keypad. The Home and End keys are located on the row of function keys, while the Down, Left and Right arrow keys are nested under the lowermost row of the text entry and numeric keypad sections of the keyboard. This allows the Insert, Delete, Page Up and Page Down keys to be disposed in a single column as opposed to three columns in the keyboard of the prior art.

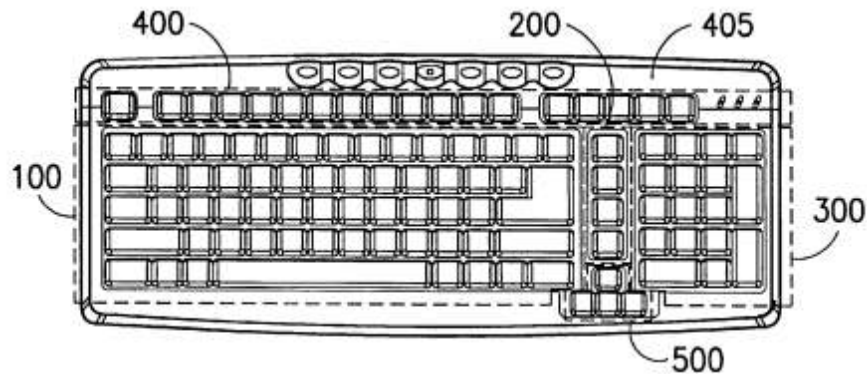

This new arrangement of single column navigation keys saves a lot of space and makes the keyboard compact. The alphanumeric keys are arranged in the conventional QWERTY format to maintain user familiarity.

**TRIZ based analysis**

The invention reorganizes the keys on the keyboard, by relocating the navigation keys and PgUp/ PgDn keys to the unused surface of the keyboard **(Principle-17: Another dimension)**.

**2.9 Keyboard for handheld computers (Patent 6507336)**

**Background problem**

The personal digital assistant (PDA) and palm-sized computers also need input devices like keyboards, stylus or touch sensitive displays. However, a stylus is more efficient for graphics input and not for entering text. Keyboards (even folding keyboards) are too large and cumbersome for handheld computers as the complete device is intended to be kept inside a pocket. Secondly a palm top is usually held with one hand and operated with another; hence an add-on keyboard is not suitable to operate.

**Solution provided by the invention**

Lunsford developed a new keyboard (patent 6507336, assigned to Palm Inc., Jan 03) to overcome the difficulties of separate physical keyboard. The invention discloses a keyboard actuated by a stylus of the handheld computer.

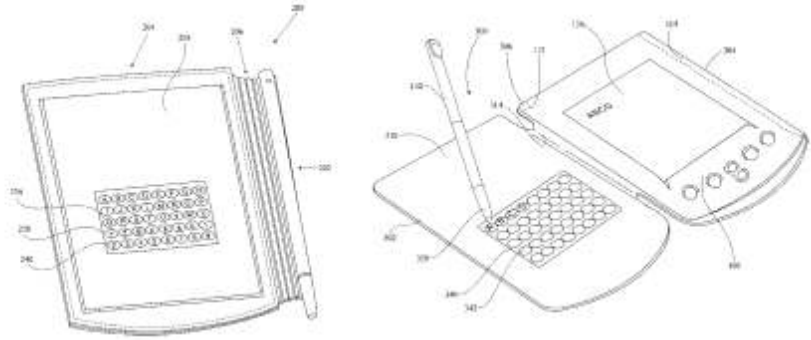

This implementation is better than a folding keyboard as it is more compact and attached to the computer. It is better than a soft keyboard or virtual keyboard displayed on the screen. A stylus-based keyboard is also convenient for physically handicapped users who can hold the stylus on mouth and operate the keyboard.

While the keys in the typical keyboard are convex for easy access by human fingers, the keys in the invention are concave for easy access by the stylus. The concave surface is kept below the surface of the encasement cover, which avoids undesirable contact or friction against the display screen.

**TRIZ based analysis**

The invented keyboard is actuated with a stylus instead of human fingers **(Principle- Mechanics Substitution)**.

Contrary to the conventional convex keys of a typical keyboard, the invention uses concave keys for easy access by a stylus **(Principle-13: Other way round)**.

**2.10 Snap-on keyboard and method of integrating keyboard (Patent 6573843)**

**Background problem**

Desktop computers are designed with detachable input/output devices such as monitors, keyboards, mice etc. that allows it to configure in the most desirable position and most ergonomic manner. In contrast, most components of the laptops are fixed. Some portable computers provide an auxiliary keyboard port to connect an external keyboard but an external keyboard covers a lot of desk space. There is a need to provide an ergonomic keyboard to portable computers without affecting its portability.

**Solution provided by the invention**

Stephen Murphy invented a detachable snap-on keyboard (Patent 6573843, Assigned to Micron Technology, June 03) for the laptops. The snap-on keyboard can be placed over the built in keyboard without needing additional table space.

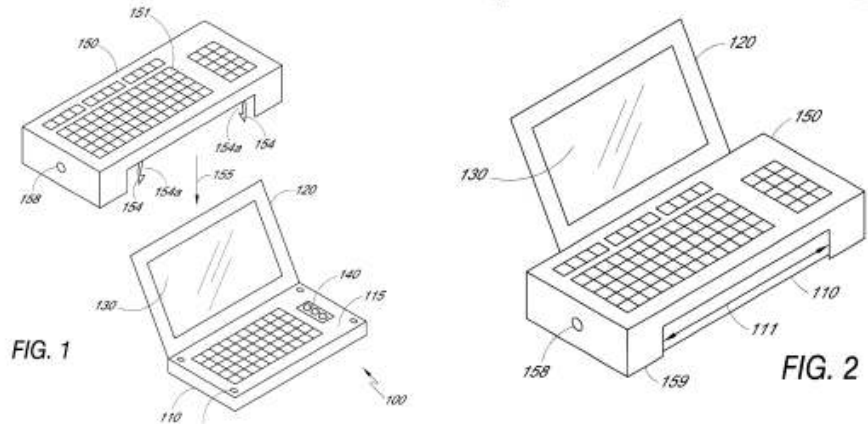

The snap-on keyboard is attached to a connector on the laptop. The connector is configured to detect the snap-on keyboard when it is kept on the surface of the laptop and pressed downward.

**TRIZ based analysis**

The invention uses a snap-on keyboard, which is placed on the laptop keyboard **(Principle-7: Nested Doll)**.

The laptop keyboard has a cavity or slot on the top to connect the external snap-on keyboard **(Principle-31: Hole)**.

# 3. Summary and conclusion

**Methods of reducing keyboard size**

Various methods are followed to reduce the size of a keyboard. Each method has some advantages and disadvantages, which have been overcome by different inventions. The following are some methods to reduce keyboard size.

- Reduce the size of the keys, so that all the keys can be accommodated in a small size keyboard.

- Reducing the gaps between keys, either horizontally or vertically or both.

- Reduce the number of keys, so that the size of the keyboard can be reduced. (The inventions on reducing number of keys are presented in a separate article).

- Eliminating duplicate keys, such as duplicate numeric keys and cursor control keys.

- Reorganizing the keyboard layout to conserve space.

- Reorganizing cursor control keys and other special keys while keeping the basic QWERTY structure in tact.

- Making a compressible keyboard that can be expanded when used and compressed when not in use.

- Make a folding structure of the keyboard so that the big keyboard can be folded in to a small place (The inventions on folding keyboard are presented in a separate article).

**Conclusion**

With the growing demand of portable computers there is an obvious demand for small size keyboards. However, this leads to a classical contradiction between the size of the keyboard and comfort of typing.

When the keyboard is meant for a small portable computer the size of the keyboard cannot be compromised for typing comfort. In most cases the user may sacrifice typing comfort to some extent to get a small size keyboard that fits well to his portable computer.

Various inventions illustrated in this article have attempted to solve this contradiction so that the user achieves both the benefits of reduced size and typing comfort.